\documentclass[twocolumn,showpacs,preprintnumbers,amsmath,amssymb]{revtex4}
\usepackage{graphicx}
\input epsf
\usepackage{dcolumn}
\usepackage{bm}

\begin{document}


\title{Brane-like singularities with no brane
}
\author{A.V. Yurov}
\email{artyom_yurov@mail.ru}
{%
\affiliation{%
I. Kant Russian State University, Theoretical Physics Department,
 Al.Nevsky St. 14, Kaliningrad 236041, Russia
\\
 }

{%
}%


\date{\today}


\begin{abstract}
 We use  a  method of linearization  to study the emergence of
the future  cosmological singularity characterized by finite value
of the cosmological radius. We uncover such singularities that  keep
Hubble parameter finite while making  all higher derivatives
of the scale factor (starting out from the $\ddot a$) diverge as the cosmological singularity is approached. Since
such singularities has been obtained before in the brane world model we
name them the ''brane-like'' singularities.  These singularities
can occur during the expanding phase in usual Friedmann universe
filled with both a self-acting, minimally coupled scalar field and
a homogeneous tachyon field.  We discover a
new type of finite-time, future  singularity which
is different from type I-IV cosmological singularities in that it
has the scale factor, pressure and density finite and nonzero. The
generalization of $w$-singularity is obtained as well.

\end{abstract}


\maketitle

\section{Introduction}

Starting out from the discovery of the cosmic acceleration \cite{1}
there have been constructed  many models of the dark energy, including the very unusual ones:
the phantom energy, the tachyon cosmologies,  the brane worlds etc. Consideration of these models results in some unexpected
conclusions about possibility of new cosmological doomsday
scenarios: the Big Rip singularity (BRS) \cite{BR}, the Big Freeze
singularity (BFS) \cite{BF}, \cite{Sahni}, the Sudden Future
singularity (SFS) \cite{SF}, the Big Boost singularity (BBtS)
\cite{boost}, and the Big Break singularity (BBS) \cite{BB},
\cite{BB1}. In all these models the evolution ends  with the
curvature singularity, $|{\ddot a}(t)|\to\infty$, reachable in a
finite proper time, say as $t\to t_s$. BRS and BFS both take place in the
phantom models but with the different equations of state. In particular, BRS takes place if $w=p/\rho={\rm const}<-1$ whereas BFS occurs for the dark energy in the form of a phantom generalized Chaplygin gas. Models with  the SFS, BFS, BBtS and BBS singularities are
characterized by a finite value of the cosmological radius but
different values of Hubble expansion parameter $H_s=H(t_s)$ and
different signs of (divergent) expression ${\ddot a_s}/a$ (cf.
also \cite{Oddd}):
$$
a_s=\infty,\qquad H_s=+\infty,\qquad \frac{\ddot
a_s}{a_s}=+\infty, \eqno({\rm BRS})
$$
$$
a_s<\infty,\qquad H_s=+\infty,\qquad \frac{\ddot
a_s}{a_s}=+\infty, \eqno({\rm BFS})
$$
$$
a_s<\infty,\qquad 0<H_s<\infty,\qquad \frac{\ddot
a_s}{a_s}=-\infty, \eqno({\rm SFS})
$$
$$
a_s<\infty,\qquad 0<H_s<\infty,\qquad \frac{\ddot
a_s}{a_s}=+\infty, \eqno({\rm BBtS})
$$
$$
a_s<\infty,\qquad H_s=0,\qquad \frac{\ddot a_s}{a_s}=-\infty.
\eqno({\rm BBS})
$$
{\it Remark 1}. One of classifications of singularities for the modified
gravity was given in \cite{Mod-grav1}; for the classification  and discussion concerned with
avoiding the singularities in the alternative gravity dark energy
models cf. \cite{alter}. Another classification of finite-time
future singularities (Type I-IV singularities) is presented in
\cite{uh-1}. According to this classification, the BRS is
a singularity of  type I, BFS is of type III, SFS and BBtS are type II and
BBS -- type II with $\rho_s\equiv \rho(t_s)=0$ (although this is
a quite non-trivial special case of a type II singularities). Our
classification doesn't contain singularities of IV type (for $t\to
t_s$, $a\to a_s$, $\rho\to 0$, $|p|\to 0$ and higher derivatives
of $H$ diverge) but as we shall see in Sec. VI, the classification
of Ref. \cite{uh-1} is not exactly complete too: the type IV is
the special case of a more general type of singularities.
\newline
{\it Remark 2}. Another type of ''singularity'' - so called
$w$-singularity was obtained in \cite{w-sing}. This
''singularity'' has a finite scale factor, vanishing energy density
and pressure, and the singular behavior manifesting itself only in
a time-dependent barotropic index $w(t)$. The $w$-singularities
seem to be most similar to the type IV but are different nonetheless since they do not lead to any divergence of higher order
derivatives of $H$ \cite{w-sing}.

One surmises that $w$-singularity is not a correct physical singularity
since all the physical values (i.e. density, pressure and higher
derivatives of the scale factor or Hubble roots) are finite.
Moreover, the definition of $w$-singularity from the \cite{w-sing}
is an incomplete one. To show this let us consider the following form of
the scale factor
\begin{equation}
a(t)=a_s-A\left(t_s-t\right)^m.
\label{w-contr}
\end{equation}
(\ref{w-contr}) is the special  case of the general form of the
scale factor from the \cite{w-sing} (with $B=0$, $A=a_s$,
$C/t_s^n=-A$, $D=1$, $n=m$). One can show that for $t\to t_s$:
\newline
(a) type III singularity if $0<m<1$;
\newline
(b) type II singularity if $1<m<2$;
\newline
(c) $w$-singularity if $m>2$.

The case $m=1$ results in a  model with the constant barotropic
index $w=-1/3$. The case $m=2$ is the most interesting one because
$$
\rho\to 0,\qquad p\to\frac{4A}{3a_s}\ne 0,\qquad |w|\to\infty.
$$
and
$$ \frac{d^{2n}H}{dt^{2n}}=0,\qquad
\frac{d^{2n+1}H}{dt^{2n+1}}\sim \frac{A^{n+1}}{a_s^{n+1}}<\infty,
$$
at $t=t_s$. Thus we have some generalization of $w$-singularity
such that the pressure is non-vanishing and finite at $t=t_s$.

 The BRS and BFS  have been obtained in the phantom
cosmologies (BRS for the phantom perfect fluid with equation of
state $p/\rho=w={\rm const}<-1$ and BFS for the  phantom Chaplygin
models. Throughout the paper  we'll stick to the metric units with
$8\pi G/3=c=1$). The BBtS is connected to the effect of the
conformal anomaly that drives the expansion of the Universe to a
maximal value of the Hubble constant, after which the solution
becomes complex. The BBS takes place in tachyon models.

Unlike BRS, BFS and BBtS altogether, the BBS and  SFS are violating just
the dominant energy condition ($\rho\ge 0$, $-\rho\le p\le \rho$).
It is also possible to obtain some generalization of these
singularities. In particular,  generalization of the Sudden Future
singularities (the so called Generalized Sudden Future singularities
or GSFS) are singularities such that one has the derivative of pressure
$p^{(m-2)}$ singularity which accompanies the blow-up of the
$m$-th derivative of the scale factor $a^{(m)}$ \cite{uh1}. These
singularities are possible in theories with higher-order curvature
quantum corrections \cite{uh-1} and corresponds to classification
in this paper.

Despite the fact that there has recently been a great inflow of articles, elaborating on the aforementioned singularities, an absolute majority of them has been of a mathematical nature, while the physical reasons for arousal of such singularities still remain less then clear. A remarkable exception is the article \cite{Sahni}, which has introduced for the first time a new type of cosmological singularities located on the brane (for discussion about the soft singularities on brane with the quantum corrections cf. \cite{soft-br}).
 These singularities are
characterized by the fact that while the  Hubble parameter and scale factor remain
finite, all higher derivatives of the scale factor (${\ddot a}$
etc.) diverge as the cosmological singularity is approached. These
singularities may be obtained as the result of embedding of
(3+1)-dimensional brane in the bulk and this is why these
singularities will be henceforth referred to as the ''brane-like''
singularities. We'll define the  ''brane-like''
singularities in a following fashion:
we'll say that {\em singularity is of a ''brane-like''
type if at the instance of its occurrence both scale factor and density remain finite and nonzero, while all  the higher order derivatives of scale factor (starting with the second order) become altogether singular, i.e. $a\to a_s$, $\rho\to \rho_s$, $0<a_s<\infty$, $0<\rho_s<\infty$, $d^na_s/dt^n=\infty$ for $n>1$.}

Evidently, the class of ''brane-like'' singularities includes the singularities of Type II (with $\rho_s\ne 0$) or SFS and BBtS. Moreover, BBS will also be of this type whenever we are talking about the models with the constant positive curvature, since at the singularity point $\rho_s=1/a_s^2$.

The physical nature of  ''brane-like'' singularities emergence is quite clear:  in the simple case with  $Z_2$ reflection symmetry  and the identical cosmological constants on the two sides of the brane, the dynamical equation contains few additional terms. One of them is the square root of the sum of contributions of  density (on the brane), tension, cosmological constant and the ''dark radiation'' (the last one  arises due to the projection of the bulk gravitational degrees of freedom onto the brane \cite{Sahni}). This sum is not positively defined and  might become negative during the cosmological evolution. Thus, the solution of the cosmological equations can't be continued beyond the point where this sum turns to zero and what we end up with at this point is nothing but a ''brane-like'' singularity. Since the existence of such singularities is natural in the  brane physics, it won't  be against the  logic to assume that the appearance of ''brane-like'' singularities in usual Friedmann cosmology (SFS or BBtS) might be an evidence of validity of the brane hypothesis. Therefore it is interesting to consider ''brane-like'' singularities without a brane (i.e. in FLRW cosmology) to establish the particular form of potential and the equation of state that will result in such singularities during cosmological dynamics. Such potential and the equation of state may altogether be useful for answering the big cosmological question: Don't we really live on the brane?


Furthermore, such singularities may actually result in very unusual
models. In fact, let's consider the universe
which contains a ''brane-like'' singularity. If the universe is filled
with a scalar field while the Hubble parameter $H(t_s)=H_s$ and the scale
factor $a(t_s)=a_s$ are finite at the singular point
($H_s<\infty$, $a_s<\infty$) then the value of the scalar field
$\phi(t_s)=\phi_s$ might be finite as well. On the other hand, quantum
corrections of higher order  (in N-loops approximation) depend
on the higher derivatives. If higher derivatives of scale factor
diverge then this will also be the case for the scalar field. So one
can expect that since all higher derivatives of  scale factor
and field alike diverge as the cosmological singularity is approached,
then the quantum effects will be dominating  for $t\to t_s$. This will
be the case in spite of the fact that both density $\rho_s$ and scale
factor $a_s$ will be finite and that $\rho_s$ might be small and
$a_s$ -- very large.

It may seem that quantum corrections will be dominating
because the pressure $|p|\to\infty$ as the cosmological
singularity is approached. This is not the case for the
singularities of the IV type being ''brane-like'' by
definition. Moreover, in Sec. VI we'll construct the singularities of even more general type
that will violate the classifications of
\cite{uh-1} since $\rho_s$ and $p_s$ will  be finite and nonzero and
all higher derivatives will diverge.

In this paper one constructs ''brane-like'' singularities in
Friedmann-Lemaitre-Robertson-Walker universe filled with the usual   self-acting, minimally
coupled scalar field or homogeneous tachyon field.
In this cosmology we'll also construct the singularities  (with finite
scale factor) where that Hubble variable vanishes and  all higher
derivatives of the scale factor diverge as the cosmological
singularity is approached. That type of singularities is the
generalization of the Big Break singularities and there will also be those of the ''brane-like'' type for the case of a constant positive curvature. We will calculate both self-acting potential $V(\phi)$ and tachyon potential $V(T)$ that result in appearance of such singularities. Additionally, we'll present  the equation of state  for such models. Moreover, the  classification of singularities from  the Ref. \cite{uh-1}
will be complemented.

This paper contains a discussion of a  simple but useful method
which allows one to construct exact  solutions of the cosmological
Friedmann equations filled with both self-acting, minimally
coupled scalar field and a homogeneous tachyon field. The method
itself will be denoted as the method of linearization and it has been previously
suggested in \cite{Temik} (see also \cite{1Chervon}, \cite{YV},
\cite{YV-1}). We'll give a brief discussion of this method in the
next Section. In Sec. III we  construct exact solutions with
''brane-like'' singularities. Then we consider field models (both
tachyon and the usual minimally coupled scalar field) near the
singularity (Sec. IV). Finally, in Sec. V we show that method of
linearization allows one to obtain exact forms of potentials in
tachyon models. In particular, we'll show that tachyon model that
has been discussed in detail in \cite{BB} is one of the simplest models
in framework of the method of linearization. In Sec. VI we
construct the new type of singularities which are some kind of generalization
of the type IV singularities. The discussion is concluded in Sec. VII.

\section{The method of linearization}

Let us write the Friedmann equations as
\begin{equation}
\left(\frac{\dot a}{a}\right)^2=\rho-\frac{k}{a^2},\qquad 2\frac{{\ddot
a}}{a}=-(\rho+3p). \label{Fried}
\end{equation}

The crucial point of this paper is the fact that a Friedmann
equations admits a linearizing substitution and can therefore be studied
via the different powerful mathematical methods which were specifically developed
for the linear differential equations. This is the reason we
call our approach the method of linearization \cite{Temik}:
\newline
{\bf Proposition.} Let $a=a(t)$ (with $p=p(t)$, $\rho=\rho(t)$) be
a solution of (\ref{Fried}). Then for the case $k=0$ the function $\psi_n\equiv
a^n$ is the solution of the Schr\"odinger equation
\begin{equation}
\frac{{\ddot \psi_n}}{\psi_n}=U_n, \label{Scr}
\end{equation}
with potential
\begin{equation}
U_n=n^2\rho-\frac{3n}{2}\left(\rho+p\right). \label{Un}
\end{equation}
For example:
$$
U_1=-\frac{\rho+3p}{2},\qquad U_2=\rho-3p,\qquad
U_3=\frac{9}{2}(\rho-p),
$$
or
$$
U_{1/2}=-\frac{1}{2}\left(\rho+\frac{3p}{2}\right),\qquad
U_{-1}=\frac{5\rho+3p}{2}
$$
and so on.
\newline
{\it Remark 3}. If the universe is filled with scalar field $\phi$
whose Lagrangian is
\begin{equation}
L=\frac{{\dot\phi}^2}{2}-V(\phi)=K-V, \label{Lagrangian}
\end{equation}
 then the expression (\ref{Un}) will be
$$
U_n=n(n-3)K+n^2V.
$$
In particular case $n=3$ $U_3=9V(\phi)$. This particular case has
been extensively studied in \cite{1Chervon}, \cite{YV}.
\newline
{\it Remark 4}. For small values of $n\ll 1$ one gets $U_n\sim
-3n(\rho+p)/2$; for example, if $n=0.01$ then
$$
U_n\sim -(0.0149\rho+0.015p)\sim -0.015(\rho+p).
$$
Therefore one can use $U_n<0$ to check whether the weak energy
condition is violated~\footnote{At the same time, one shall keep
in mind that this equation is nothing but approximate. To ensure
(with the help of $\psi_n$) that the weak energy condition will
indeed be violated, one can use exact equation
${\dot\sigma_n}=-3n(\rho+p)/2$, where
$\sigma_n={\dot\psi_n}/\psi_n$.}. If, on the contrary, $n\gg 1$
then $U_n\sim n^2\rho$.
\newline
{\it Remark 5}. If $k=\pm 1$ then the Proposition is valid only for the case $n=0,\,1$.

As we  have shown in \cite{YV-1}, the representation  of the
Einstein-Friedmann equations as a second-order linear differential
equation (\ref{Scr}) allows for a usage of an arbitrary (known) solution for
construction of another, more general solution parameterized by a set of $3N$
constants, where $N$ is an arbitrary natural number. The large
number of free parameters should prove itself useful for constructing a
theoretical model that agrees satisfactorily with the results of
astronomical observations. In particular, $N=3$  solutions in the
general case already exhibit inflationary regimes \cite{YV-1}. Unlike the previously studied two-parameter solutions (see
\cite{1Chervon}, \cite{YV}), these three-parameter solutions might
describe an exit from inflation without any fine tuning of parameters as well as the several consecutive inflationary regimes.

In the next Section we will show that the method of linearization
is indeed an effective  one for construction of a ''brane-like'' singularity.

\section{''Brane-like'' singularity}

Assume
\begin{equation}
U(t)=\frac{\kappa u_s^2}{\left(t_s-t\right)^{\alpha}},
\label{poten}
\end{equation}
with $u_s^2={\rm const}>0$, $\kappa=\pm 1$, $\alpha>0$. For
simplicity one has omitted the index $n$: $U_n\to U(t)$, $\psi_n\to
\psi$. For $|t-t_s|\gg 1$  the potential $U(t)\to 0$ so
$\psi(t)\sim t$ and $a(t)\sim t^{1/n}$. If $n=2$ we have a
universe filled with radiation ($w=1/3$), and for $n=3/2$ we
have a dust universe with $w=0$.

Now let us consider the solution of the (\ref{Scr}) at $t\to t_s$:
\begin{equation}
\psi(t)=\psi_s+\sum_{k=1}^{\infty}
c_k\left(t_s-t\right)^{k(2-\alpha)}, \label{solution-I}
\end{equation}
where
\begin{equation}
\psi_s=a_s^{1/n}=\frac{(2-\alpha)(\alpha-1)}{\kappa u_s^2},
\label{pf-I}
\end{equation}
and
\begin{equation}
c_k=\frac{\kappa^{k-1}u_s^{2(k-1)}}{(2-\alpha)^{k-1}k!\prod_{m=0}^{k-2}\left[(k-m)(2-\alpha)-1\right]},
\label{ck}
\end{equation}
with $c_1=1$. For $\alpha <2$ the series (\ref{solution-I}) is
convergent for any $t$ (including $t=t_s$) since its radius is
$$
R=\lim_{k\to\infty}\mid\frac{c_k}{c_{k+1}}\mid=+\infty.
$$
If $\alpha>2$ then the first two terms of (\ref{solution-I}) are
$$
\psi=\psi_s+\frac{1}{(t_s-t)^{\alpha-2}},
$$
so the function $\psi(t)$ will be singular at $t\to t_s$ and
we have either Big Rip ($n>0$) or Big Crunch ($n<0$) at $t=t_s$.  If
$\alpha=2$ the general solution of the (\ref{Scr}) will be of the form
\begin{equation}
\psi=\sqrt{t_s-t}\left(C_1(t_s-t)^L+C_2(t_s-t)^{-L}\right),
\label{a2}
\end{equation}
with the arbitrary constants $C_{1,2}$ and $L=\sqrt{1+4\kappa
u_s^2}/2$. For any $C_1$ and $C_2$ the solution (\ref{a2}) results
in Big Rip or Big Crunch singularity as well. Finally, if $\alpha=1$
then the series (\ref{solution-I}) will be
$$
\begin{array}{l}
\displaystyle{ \psi=\left(t_s-t\right)[1+\frac{\kappa
u_s^2}{2}(t_s-t)+\frac{u_s^4}{12}(t_s-t)^2+}\\
\\
\displaystyle{ +\frac{\kappa
u_s^6}{144}(t_s-t)^3+\frac{u_s^8}{2880}(t_s-t)^4+...]}
\end{array},
$$
so there is no cosmological singularity and $a$ is finite.

Thus, in framework of our investigation one must consider only
$0<\alpha<1$ and $1<\alpha<2$. Keeping only the two first terms in
(\ref{solution-I}) and using
\begin{equation}
H=\frac{\dot\psi}{n\psi},\qquad \frac{\ddot
a}{a}=\frac{\ddot\psi}{n\psi}-(n-1)H^2, \label{formuly}
\end{equation}
one gets
\begin{equation}
\begin{array}{l}
\displaystyle{ \psi\sim \frac{(2-\alpha)(\alpha-1)}{\kappa
u_s^2}-\left(t_s-t\right)^{2-\alpha},}\\
\\
\displaystyle{ H=\frac{\kappa
u_s^2}{n(\alpha-1)}\left(t_s-t\right)^{1-\alpha},}\\
\\
\displaystyle{ \frac{\ddot
a}{a}=\frac{u_s^2}{n(t_s-t)^{\alpha}}\left[\kappa-\frac{u_s^2(n-1)}{n(\alpha-1)^2}\left(t_s-t\right)^{2-\alpha}\right].}
\label{uh-ty-kakoj}
\end{array}
\end{equation}
Using (\ref{uh-ty-kakoj}) one can show that the conditions
$\psi_s>0$ and $H>0$ will hold if $n>0$. So one has two cases:

(i) If $\kappa=-1$ then $0<\alpha<1$ and one gets BBS;

(ii) if $\kappa=+1$ then $1<\alpha<2$ and one gets BFS.

To obtain SFS one should use another solution of (\ref{Scr}):
\begin{equation}
\begin{array}{l}
\displaystyle{ \psi=\psi_s-n\psi_sH_s(t_s-t)+\frac{\kappa
u_s^2\psi_s}{(2-\alpha)(1-\alpha)}\left(t_s-t\right)^{2-\alpha}+}\\
\\
\displaystyle{ \sum_{k=1}^{\infty}
c_{2k}\left(t_s-t\right)^{(2-\alpha)(k+1)}+ \sum_{k=0}^{\infty}
c_{2k+1}\left(t_s-t\right)^{(2-\alpha)(k+1)+1},}
\label{solution-II}
\end{array}
\end{equation}
where $H_s={\rm const}>0$, $\psi_s=a_s^{1/n}={\rm const}>0$ and
$$
\displaystyle{ c_{2k}=\frac{(\kappa
u_s^2)^{k+1}\psi_s}{(2-\alpha)^{k+1}(k+1)!
\prod_{m=1}^{k+1}\left[m(2-\alpha)-1\right]},}
$$
$$
\displaystyle{ c_{2k+1}=-\frac{(\kappa
u_s^2)^{k+1}nH_s\psi_s}{(2-\alpha)^{k+1}(k+1)!
\prod_{m=1}^{k+1}\left[m(2-\alpha)+1\right]}.}
$$
Using (\ref{formuly}) we get
$$
H(t_s)=H_s,\qquad \left(\frac{\ddot a}{a}\right)_{t\to
t_s}=\frac{\kappa u_s^2}{n(t_s-t)^{\alpha}}-(n-1)H_s^2,
$$
so the solution (\ref{solution-II}) contains the SFS at $t\to t_s$
for $\alpha<1$.

At last, let's consider the  (\ref{solution-I}). After
differentiation one gets
\begin{equation}
\frac{d^m\psi}{dt^m}=(-1)^m\sum_{k=1}^{\infty}
c_k\prod_{l=0}^{m-1}\left[k(2-\alpha)-l\right]\left(t_s-t\right)^{k(2-\alpha)-m},
\label{proiz}
\end{equation}
so we have a singularity in (\ref{proiz}) for
\begin{equation}
k<\frac{m}{2-\alpha}. \label{nerav}
\end{equation}
Since $0<\alpha<1$ then
$$
\frac{1}{2}<\frac{1}{2-\alpha}<1.
$$
Therefore for  $m>1$ the expression (\ref{proiz}) diverge as
the cosmological singularity is approached. Thus we have obtained
the solution which contains a some kind of generalization of the Big Break
singularity (the scale factor remains finite and the Hubble parameter
vanishes as  singularity is approached). In the case of positive curvature one has to choose $n=1$ (cf. Remark 5) and we have a
''brane-like'' singularity (the density is finite and positive whereas all higher derivatives of the
scale factor, starting out from the second one, diverge as the
cosmological singularity is approached). In the next Section we'll present a couple of models with  the self-acting and
minimally coupled scalar fields or with the homogeneous tachyon fields
$T=T(t)$ described by Sen’s or Born-Infeld type Lagrangians which result in such behavior.

The similar investigation can be done for the (\ref{solution-II}).
 This singularity is characterized by the fact that Hubble parameter  remains
finite instead of vanishing as the cosmological singularity is
approached.  This solution describes the appearance of a ''brane-like'' singularity in the flat universe.

 \section{Field models}

If the universe is filled with a self-acting  and minimally
coupled scalar field with Lagrangian (\ref{Lagrangian}) then the
energy density and pressure are
$$
\rho=K+V,\qquad p=K-V,
$$
therefore
\begin{equation}
V=\frac{1}{2}(\rho-p),\qquad K=\frac{1}{2}(\rho+p). \label{Vphi}
\end{equation}
Using (\ref{Scr}) one can write
\begin{equation}
K=\frac{\dot\psi^2-\psi\ddot\psi}{3n\psi^2}=\frac{(w+1)\dot\psi^2}{2n^2\psi^2},
\label{Kw}
\end{equation}
\begin{equation}
V=\frac{n\psi\ddot\psi+(3-n)\dot\psi^2}{3n^2\psi^2}=\frac{(1-w)\dot\psi^2}{2n^2\psi^2},
\label{Vw}
\end{equation}
where
\begin{equation}
w=\frac{p}{\rho}=-1+\frac{2n}{3}\left(1-\frac{\ddot\psi\psi}{\dot\psi^2}\right).
\label{ww}
\end{equation}
The second model is the universe filled with a homogeneous tachyon
field $T=T(t)$ described by the Sen’s Lagrangian density \cite{Sen}:
\begin{equation}
L=-V(T)\sqrt{1-g_{00}{\dot T}^2}, \label{LT}
\end{equation}
and in a flat Friedmann universe with metric
$ds^2=dt^2-a^2(t)dr^2$ we have density and pressure
\begin{equation}
\rho=\frac{V(T)}{\sqrt{1-{\dot T}^2}}, \qquad p=L, \label{rpT}
\end{equation}
so
\begin{equation}
{\dot T}^2=1+w,\qquad V=\frac{{\dot \psi}^2}{n^2\psi^2}\sqrt{-w},
\label{TV}
\end{equation}
where $w=p/\rho$ is defined by the (\ref{ww}).

The expression for $V$  (\ref{TV}) holds iff $w<0$. If $w>0$
one should introduce a new field theory based on a
Born-Infeld type action with Lagrangian \cite{BB}
\begin{equation}
L=W(T)\sqrt{{\dot T}^2-1}, \label{LT-1}
\end{equation}
so
$$
\rho=\frac{W(T)}{\sqrt{\dot T^2-1}},\qquad p=L,
$$
and
\begin{equation}
{\dot T}^2=1+w,\qquad W=\frac{{\dot \psi}^2}{n^2\psi^2}\sqrt{w}.
\label{TV-1}
\end{equation}
It is interesting  that models (\ref{LT}) and (\ref{LT-1}) can be
connected via the so called ''transgression of the boundaries''
\cite{BB}.

Now, using the potential (\ref{poten}) and the solution
(\ref{solution-I}) we get
\begin{equation}
w=-\frac{2n(\alpha-1)^2}{3\kappa u_s^2(t_s-t)^{2-\alpha}}.
\label{w-1}
\end{equation}
Therefore, for the case $\kappa=+1$  we have $w<0$ (BFS) and
$w\to-\infty$ as the cosmological singularity is approached. Using
(\ref{Kw}) and  (\ref{TV}) one concludes that this will be the case
for the phantom (both usual $\phi$ and tachyon $T$) fields only -
the case  which we leave out of consideration in this paper.

For the case $\kappa=-1$ and $0<\alpha<1$ one has a Big Break
singularity with $w\to +\infty$. For the model (\ref{Lagrangian})
we get
$$
\dot\phi^2=\frac{2u_s^2}{3n(t_s-t)^{\alpha}}>0,
$$
and
$$
p\to\frac{2u_s^2}{3n(t_s-t)^{\alpha}}\to+\infty,\qquad H\to 0,
$$
as the cosmological singularity is approached. So all energy
conditions are satisfied and $H_s=0$ as it should be for a BBS. If $k=+1$ we have a ''brane-like'' singularity with
\begin{equation}
\rho_s=\frac{u_s^4}{(2-\alpha)^2(1-\alpha)^2}.
\label{RRRS}
\end{equation}
 At $t\to t_s$ the potential $V=V(\phi)$ and field $\phi=\phi(t)$ are
given by expressions
\begin{equation}
V(\phi)=-\frac{u_s^2}{3n}\left(\frac{3n(2-\alpha)}{8u_s}\right)^{-\alpha/(2-\alpha)}
\left(\phi-\phi_s\right)^{-2\alpha/(2-\alpha)},
\label{Vfi}
\end{equation}
$$\phi=\phi_s\mp\frac{2u_s}{2-\alpha}\sqrt{\frac{2}{3n}}\left(t_s-t\right)^{1-\alpha/2},
$$
and $V(\phi)\to -\infty$, $\phi\to\phi_s$ at $t\to t_s$.

For the solution (\ref{solution-II}) we have the same potential (\ref{Vfi}) and
$$
w=-\frac{2\kappa u_s^2}{3nH_s^2(t_s-t)^{\alpha}},
$$
instead of (\ref{w-1}).

In the case of tachyon cosmology one should use the model
(\ref{LT-1}). It easy to see that $\dot T^2>0$ and
$$
T(t)=T_s\mp 2\sqrt{\frac{2n}{3}}\frac{1-\alpha}{\alpha
u_s}\left(t_s-t\right)^{\alpha/2}\to T_s,
$$
\begin{equation}
W(\Phi)=\sqrt{\frac{2}{3n}}\frac{u_s^3}{n(1-\alpha)}
\Phi^{2/\alpha-3}\to 0,
\label{VTh}
\end{equation}
with
$$
\Phi=\frac{1}{2}\sqrt{\frac{3}{2n}}\frac{\alpha
u_s}{1-\alpha}\left(T-T_s\right).
$$
Since $0<\alpha<1$ then $2/\alpha-3>-1$. For example, if
$\alpha=2/5$ then
$$
W(T)=\frac{5\sqrt{6} u_s^5}{54 n^{5/2}}\left(T-T_s\right)^2,
$$
and for the $\alpha=2/7$
$$
W(T)=\frac{21\sqrt{6} u_s^7}{12500 n^{7/2}}\left(T-T_s\right)^4.
$$
Finally, let us consider the equation of state which results in the potential (\ref{Vfi}). It was shown in \cite{BB} that the equation of state
\begin{equation}
\rho+p=\gamma\rho^{\lambda},
\label{EoS-1}
\end{equation}
results in dynamics which might be described by the self-acting potential is the form
\begin{equation}
V(\phi)=Q^{-2/(\lambda-1)}-\frac{\gamma}{2}Q^{-2\lambda/(\lambda-1)},
\label{poten-Kam}
\end{equation}
with
$$
Q=\frac{3\sqrt{\gamma}(\lambda-1)(\phi-\phi_s)}{2},
$$
where $\gamma>0$, $\lambda>1$. When $\phi\to\phi_s$ we have
$$
V(\phi)\to -\frac{\gamma}{2}Q^{-2\lambda/(\lambda-1)}\to-\infty.
$$
Unfortunately, this expression is just formally equivalent to  potential (\ref{Vfi}). In fact, the second term in (\ref{poten-Kam}) (which is the dominant one at $\phi\to\phi_s$) is exactly (\ref{Vfi}) if $\alpha=2\lambda/(2\lambda-1)$, therefore for $\lambda>1$ we get $1<\alpha<2$. It means that $\rho_s=\infty$ and we have a singularity of the III type which, in the case of general position, is not  a ''brane-like'' one.

Near the singularity, the correct equation of state for the solution (\ref{solution-I}) in case of a positive curvature has the form which looks similar to (\ref{EoS-1}):
\begin{equation}
\rho+3p=\gamma\left(\rho_s-\rho\right)^{-|\lambda|},
\label{EoS-2}
\end{equation}
where
$$
|\lambda|=\frac{\alpha}{2(1-\alpha)},
$$
and $0<\alpha<1$.

\section{Tachyon potentials}

The method of linearization proves to be extremely useful in finding the
potentials of the exact solvable tachyon models. In particular, as we
shall see, the tachyon model which was discussed in detail in
\cite{BB} is one of the simplest models in framework of the method
of linearization.

Let's consider Eq. (\ref{Scr}) with potential $U=0$. The
solution of this equation $\psi=Ct+C'\to Ct$ by the translation
$t\to t-C'/C$. In this case we get
$$
w=\frac{p}{\rho}=-1+\frac{2n}{3},\qquad
p=\frac{2n-3}{3n^2t^2},\qquad {\dot T}^2=w+1,
$$
so
$$
T=\pm\sqrt{\frac{2n}{3}}\left(t-t_s\right)+T_s,
$$
and
\begin{equation}
V(T)=\frac{2\sqrt{9-6n}}{n\left[2(T-T_s)\pm\sqrt{6n}
t_s\right]^2}. \label{pr-1}
\end{equation}
If $n=3(1+k)/2$ with $-1<k<+1$ (so $0<n<3$) and $t_s=0$ then
(\ref{pr-1}) has exactly the form of one of the potentials  from the first paper
\cite{BB} (with $T_0\to T_s$).

The case with $U=\mu^2={\rm const}>0$ is a more interesting example .
The solution of the (\ref{Scr}) with Big Bang singularity at $t=0$
is $\psi=C\sinh(\mu t)$. So
$$
w=-1+\frac{2n}{3\cosh^2\mu t},\qquad p=\frac{\mu^2(2n-3\cosh^2\mu
t)}{2n^2 \sinh^2 \mu t}.
$$
Using (\ref{TV}) one gets
$$
T(t)=\pm \sqrt{\frac{2n}{3\mu^2}} \arctan (\sinh \mu t)+T_0,
$$
and
$$
V(t)=\frac{\mu^2\cosh\mu t}{3n^2\sinh^2\mu t}\sqrt{9\cosh^2\mu
t-6n}.
$$
Introducing $\Lambda=\mu^2/n^2$, $k=2n/3-1$ we get
\begin{equation}
V(T)=\frac{\Lambda}{\sin^2\xi}\sqrt{1-(1+k)\cos^2\xi}, \label{ppp}
\end{equation}
where
$$
\xi=\frac{3}{2}\sqrt{\Lambda(1+k)} T.
$$
(\ref{ppp}) is the basic tachyon  model of the paper \cite{BB}.

It is not difficult to construct many other integrable tachyon
models using the simple, solvable  potentials of (\ref{Scr}).
Another fruitful way of doing it lies in a use of the Darboux transformation to
those initial potentials ($U=0$, $U=\mu^2$). This, however, is out of scope of the paper.

\section{Generalized type IV singularities}

The singularities of type IV (according to the classification of
Ref. \cite{uh-1}) have the following behavior: for $t\to t_s$,
$a\to a_s$, $\rho\to 0$, $|p|\to 0$ and the higher derivatives of $H$
diverge ($0<a_s<\infty$). In this section we present a new type of
singularities:
\newline
\newline
{\bf Generalized type IV}: For $t\to t_s$, $a\to a_s$, $\rho\to
\rho_s$, $p\to p_s$ and higher derivatives (starting out from the
third one) of $H$ diverge and $0<a_s<\infty$, $0<\rho_s<\infty$,
$0<|p_s|<\infty$.

Let's put $n=3$, $\psi=a^3$ and (in parametric form)
\begin{equation}
\begin{array}{l}
\psi=A+\frac{\kappa
B}{4}\left(4\cos\,\eta-2\cos^2\eta-\cos^4\eta-4\log(1+\cos\,\eta)\right),
\\
\\
\displaystyle{t=t_s+\frac{1}{\kappa}\left(\log\left|\tan\frac{\eta}{2}\right|+\cos\eta\right)},
\end{array}
\label{newsolution}
\end{equation}
where $A$, $B$, $\kappa$ are constants, $0\le\eta\le \pi$;
 $\eta=0$ corresponds to
$t=-\infty$, $\eta=\pi/2$ to $t=t_s$ (singularity) and $\eta=\pi$
to $t=+\infty$. After the differentiation we get (a dot denotes
the derivative with respect to cosmic time $t$ rather than to
parameter $\eta$)
$$
{\dot\psi}=\kappa^2 B\left(1-\cos^3\eta\right),
$$
$$
{\ddot\psi}=3\kappa^3 B\sin^2\eta,
$$
$$
{\dddot\psi}=\frac{6B\kappa^4\sin^2\eta}{\cos\eta},
$$
$$
{\ddddot\psi}=\frac{6\kappa^5B\sin^2\eta(\cos^2\eta+1)}{\cos^4\eta},
$$
and so on.

For $\kappa B>0$ the function (\ref{newsolution}) is the monotonously
increasing one for $0\le\eta\le \pi$ and $\psi(\pi)=+\infty$ (i.e.
at $t=+\infty$). Thus at $t=-\infty$
$$
\psi=A+B\kappa\left(\frac{1}{4}-\log 2\right),\qquad
\dot\psi=0,\qquad \ddot\psi=0.
$$
At $t=t_s$
$$
\psi=A,\qquad \dot\psi=\kappa^2B, \qquad \ddot\psi=3\kappa^3B
$$
and, starting out from the third one, all higher derivatives
diverge. At $t=+\infty$
$$
\psi={\rm sign}(\kappa B)\times\infty,\qquad
\dot\psi=2\kappa^2B,\qquad \ddot\psi=0.
$$
Therefore at $t=t_s$
$$
\rho_s=\frac{\kappa^4B^2}{9A^2},\qquad p_s=\frac{B\kappa^3(\kappa
B-6A)}{9A^2},\qquad w_s=1-\frac{6A}{\kappa B}.
$$
Thus we have a generalization of a type IV singularity at $t_s$,
where the density and pressure are finite and nonzero whereas all
higher derivatives  of $H$ diverge.

It is convenient to introduce a new parameter $s$:
$$
s=\frac{6A}{\kappa B},
$$
so
$$
\rho_s=\frac{4\kappa^2}{s^2},\qquad
p_s=\frac{4\kappa^2(1-s)}{s^2},\qquad w_s=1-s.
$$

If $4/3<s<2$ then $-1<w_s<-1/3$; if $s=2$ then $w_s=-1$; if $s>2$
then $w_s<-1$. To obtain the initial Big Bang singularity at
$t=t_i$, $-\infty<t_i<t_s$ one should put
$$
s<s_i=6\log 2-\frac{3}{2}\sim 2.659,
$$
or
$$
w_s>w_i=-1.659.
$$
In the initial Big Bang singularity, the barotropic index $w=+\infty$,
on the other hand, $w(\eta)$ is the monotonously decreasing function.
These properties result in a following conclusion: if
$$
\frac{4}{3}<s<s_i,
$$
then after Big Bang the model (\ref{newsolution}) will go through the usual expansion with damping, but starting out from some
moment it will experience an accelerated expansion up to a future generalized
type IV singularity.

\section{Conclusion}

In this paper we have discussed a simple  method of construction
of exact solutions of the Friedmann equations with finite scale factor
singularities. Despite simplicity, the method allows for
acquirement of solutions characterized by the extremely interesting
properties.

The main results of this work are:

(i) we have obtained a new type of finite-time, future singularities
which seem to be most similar to the type IV of \cite{uh-1} but
are different nevertheless as they have nonzero pressure
and density at the singular point;

(ii) we have obtained a new type of finite-time, future
quasi-singularities being rather similar  to the
$w$-singularities (which are quasi-singularities too) but having nonzero pressure
at a singular point;

(iii) we have shown that ''brane-like'' singularities can occur in
a common Friedmann cosmology with potential (\ref{Vfi}) and equation of state (\ref{EoS-2})
(near of singularity) as well as in a tachyon cosmology with potential (\ref{VTh});

(iv) we have obtained the generalized Big Break singularities not
only for the universe filled with tachyons but also a usual minimally
coupled scalar field;

(v) we have shown that basic tachyon model which was discussed in
detail in \cite{BB} is one of the simplest models in framework of
the method of linearization.

\acknowledgements

\noindent I'd like  to thank  A. Yu. Kamenshchik, S.D. Odintsov,
S.V. Chervon,   M.P. Dabrowski and V.A. Yurov for important notices. This work
was supported in part by the Russian Foundation for Basic Research
(Grant No. 08-02-91307-${\rm IND}_{\rm a}$ and No.
  09-05-00446a).
I would also like to express my gratitude to the European Science Foundation (ESF) for the financial support I received while participating in the project ''New Trends and Applications of the Casimir Effect'' (Grant FIS2006-02842). Additionally it is my please to thank prof. S.D. Odintsov
and prof. Dr. Emilio Elizalde for hospitality in Instituto de Ciencias del Espasio del Consejo Superior de Investigasiones Cientificas (ICE, CSIC-IEEC), en Bellaterra (Barcelona).  Finally I would also like to thank the anonymous  Referee who's critical questions helped me in my task of improving this article.


\begin{references}



\bibitem{1} A. G. Riess et al. [Supernova Search Team Collaboration],
Astron. J. {\bf 116}, 1009 (1998) [arXiv:astro-ph/9805201]; S.
Perlmutter et al. [Supernova Cosmology Project Collaboration],
Astrophys. J. {\bf 517}, 565 (1999) [arXiv:astro-ph/9812133].

\bibitem{BR} A.A. Starobinsky, Grav. Cosmol. {\bf 6}, 157 (2000)
[arXiv:astro-ph/9912054]; R.R. Caldwell, Phys. Lett. B545 (2002)
23 [arXiv:astro-ph/9908168]; R.R. Caldwell, M. Kamionkowski and
N.N. Weinberg, Phys. Rev. Lett. 91 (2003) 071301
[arXiv:astro-ph/0302506]; P.F. Gonz\'{a}lez-D\'{i}az, Phys. Lett.
B586 (2004) 1 [arXiv:astro-ph/0312579]; P.F.
Gonz\'{a}lez-D\'{i}az, Phys.Rev.Lett. 93 (2004) 071301
[astro-ph/0404045]; P.F. Gonz\'{a}lez-D\'{i}az, Phys.Rev. {\bf
D69} (2004) 103512 [astro-ph/0311244]; S.M. Carroll, M. Hoffman
and M. Trodden, Phys. Rev. D68 (2003) 023509
[arXiv:astro-ph/0301273]; S. Nojiri and S.D. Odintsov, Phys. Rev.
D70 (2004) 103522 [arXiv:hep-th/0408170]; A.V. Yurov, P.M. Moruno,
P.F. Gonzalez-Diaz, Nucl.Phys. {\bf B759} (2006) 320
[arXiv:astro-ph/0606529].

\bibitem{BF}
 M. Bouhmadi-Lopez, P. F. Gonzalez-Diaz and P.
Martin-Moruno, arXiv:gr-qc/0612135; M. Bouhmadi-Lopez, P. F.
Gonzalez-Diaz and P. Martin-Moruno, arXiv:0707.2390 [gr-qc]; A. V.
Yurov, A. V. Astashenok and P.F. Gonz\'{a}lez-D\'{i}az, Grav.
Cosmol. {\bf 14} No.3, 205 (2008), arXiv:0705.4108 [astro-ph]; F.
Cannata, A. Yu. Kamenshchik and  D. Regoli, Phys. Lett. {\bf B670}
(2009) 241, arXiv:0801.2348 [gr-qc].

\bibitem{Sahni} Y. Shtanov and V. Sahni, Class.Quant.Grav. {\bf 19} (2002)
L101 [arXiv:gr-qc/0204040].


\bibitem{SF}
 J.D. Barrow, Class. Quant. Grav. 21 (2004)
L79 [arXiv:gr-qc/0403084]; J.D. Barrow, Class. Quant. Grav. 21
(2004) 5619 [arXiv: gr-qc/0409062]; J.D. Barrow, Class. Quant.
Grav. 22 (2005) 1563 [arXiv: gr-qc/0411045]; M.P. Dabrowski,
Phys.Rev. {\bf D71} (2005) 103505 [arXiv:gr-qc/0410033]; M.P.
Dabrowski, Phys.Lett. {\bf B625} (2005) 184 [arXiv:gr-qc/0505069];
J. D. Barrow and S.Z. Lip, arXiv:0901.1626 [gr-qc].


\bibitem{boost} A.O. Barvinsky, C. Deffayet and
A. Y. Kamenshchik, JCAP 0805 (2008) 020, arXiv:0801.2063 [hep-th].



\bibitem{BB}
V. Corini, A. Yu. Kamenshchik, U. Moschella and V. Pasquier, Phys.
Rev. {\bf D69} (2004) 123512 [arXiv:hep-th/0311111]; Z. Keresztes,
L. A. Gergely, V. Gorini, U. Moschella, A. Yu. Kamenshchik, Phys.
Rev. {\bf D79} (2009) 083504, arXiv:0901.2292 [gr-qc].
\bibitem{BB1}
A. Kamenshchik, C. Kiefer and B. Sandhoefer, Phys.Rev. {\bf D76}
(2007) 064032, arXiv:0705.1688 [gr-qc].

\bibitem{Oddd} S. Nojiri and S.D. Odintsov, arXiv:0903.5231 [hep-th];
S. Nojiri and S.D. Odintsov, Phys.Rev. {\bf D72} (2005) 023003
[arXiv: hep-th/0505215].


\bibitem{Mod-grav1}
S. Nojiri, S. D. Odintsov  Phys.Rev. {\bf D78} (2008) 046006,
arXiv:0804.3519 [hep-th]; K. Bamba, S. Nojiri and S D Odintsov,
JCAP0810:045 (2008), arXiv:0807.2575 [hep-th].
\bibitem{alter} S. Capozziello, M. De Laurentis, S. Nojiri, S.D. Odintsov, arXiv:0903.2753 [hep-th]

\bibitem{uh-1} S. Nojiri, S.D. Odintsov and S. Tsujikawa, Phys. Rev.
{\bf D71}  (2005) 063004 [arXiv:hep-th/0501025].


\bibitem{w-sing} M.P. Dabrowski and T. Denkiewicz, Phys. Rev. {\bf D79} (2009)
063521, arXiv:0902.3107 [gr-qc].


\bibitem{uh1} M.P. Dabrowski, T. Denkiewicz and  M.A. Hendry, Phys.Rev. {\bf D75}
(2007) 123524, arXiv:0704.1383 [astro-ph].

\bibitem{soft-br} S Nojiri and S D Odintsov, Phys. Lett. {\bf B595}
(2004) 1 [arXiv:hep-th/0405078].


\bibitem{Temik} A.V. Yurov, A.V. Astashenok, V.A. Yurov, Grav. Cosmol., {\bf 14}
(2008) 8 [arXiv:astro-ph/0701597].

\bibitem{1Chervon} S.V. Chervon and V.M. Zhuravlev, [arXiv:gr-qc/9907051];
V. M. Zhuravlev, S. V. Chervon, and V. K. Shchigolev, JETP {\bf
87}  (1998) 223.

\bibitem{YV} A.V. Yurov, [astro-ph/0305019]; A.V. Yurov and S.D.
Vereshchagin, Theor. Math. Phys. {\bf 139}  (2004) 787
[arXiv:hep-th/0502099].

\bibitem{YV-1} A. V. Yurov and A. V. Astashenok, Theor.
Math. Phys. {\bf 158} (2009) 261, arXiv:0902.1979 [astro-ph.CO].


\bibitem{Sen} A. Sen,  JHEP 0204 (2002) 048
[arXiv:hep-th/0203211]; A. Sen,  JHEP 0207 (2002) 065
[arXiv:hep-th/0203265].




\end{references}
\end{document}